# Blockchain-based E-commerce: It's an Evolution, NOT a Revolution -- Experimental Evidence from Users' Perspective

David Lee Kuo Chuen[a], Yang Li[a], Weibiao Xu[a], Willy Zhao[b]

[a]School of Business, Singapore University of Social Sciences

[b]Department of Telematics, Norwegian University of Science and Technology

## Abstract

Proponents of blockchains believe that this technology will revolutionize e-commerce. To evaluate this belief, we invite several groups of students to transact on a decentralized peer-to-peer marketplace built on the platform provided by Origin Protocol Inc., and then we conduct a survey about their experience of usage. Based on our survey results, we find that 33% of respondents play tricks on others, which implies that this undesirable result may hinder the widespread adoption of blockchain technologies. We also attempt to propose a conceptual mechanism to mitigate fraudulent behaviors. In the event of disputation, a trusted authority is entitled to the right to downgrade the fraudulent side's credit record, which is stored by a permissioned blockchain accessed only by the authority. Such a punishment can effectively decrease agents' incentives to sell counterfeits and leave fake ratings. In sum, we must distinguish what we proposed blockchains will do and what blockchains can do before enabling this technology in e-commerce.

# 1. Introduction

E-commerce selling has been around for decades and is still growing in an undeniable trend. According to statistics, online revenue increased from $2.3 to over $2.8 trillion from 2017 to 2018 and is expected to grow to 4.88 trillion US dollars in 2021[1]. Meanwhile, the growing e-commerce market experiences trust issues and privacy leakage. For example, the Pew Research Center reported that consumers' lack of trust in online reviews presents one of the biggest problems for the growth of e-commerce.[2] Another concern is the complaints of counterfeits induced by the lack of transparency in the supply chain. In addition, online marketplaces such as Amazon, eBay, and Alibaba, despite their popularity, are often criticized for being centralized, i.e., controlled by a single entity that manages users' private information and transition details, which increases the risk of privacy leakage.

Proponents of blockchains claim that this technology can resolve the above problems by creating more advanced and decentralized online platforms. It is emphasized that centralized entities are prone to error and commercial bias; in contrast, blockchains enable a total reliance on objective code, referred to as trustlessness—a term that denotes the ability to confirm the truth without resorting to a trusted third party. With the emergence of blockchains, self-enforcing programs, such as "smart contracts," appear to be the perfect technology to enable decentralized marketplaces by eliminating counterparty risk without reliance on intermediaries.[3] Seemly, then, the widespread belief that "decentralization" or "trustlessness" regarded as characteristics of the so-called superior technologies will revolutionize e-commerce is credible.

---

[1] Figures are from https://www.statista.com/statistics/379046/worldwide-retail-e-commerce-sales/.

[2] The survey finds that in the US 49% of customers do not trust shopping online out of which, 60% of the respondents believe security and privacy policies are the primary factors; over 80% of adults in the United States occasionally refer to online customer reviews before purchasing; ironically, only half of them trust the accuracy of the product description as online reviews and ratings can be manipulated easily. https://www.pewinternet.org/2016/12/19/online-shopping-and-e-commerce/.

[3] See generally: Szabo, N., 1994. Smart contracts. http://szabo.best.vwh.net/smart.contracts.html/.

To verify the adoption of blockchains in e-commerce, we build a decentralized peer-to-peer marketplace through the platform provided by Origin Protocol Inc. and invite several groups of students to transact on it. Then, we survey their experience with using blockchain.[4] (See details of the platform and experiment in Section 3). Specifically, we conduct five trials, 4 in Tianjin, China, and 1 in Trondheim, Norway.[5] Almost 133 participants first used blockchain applications, i.e., decentralized platforms and cryptocurrencies.[6] Surprisingly, we find that 33% of subjects played tricks on others during the trading game based on feedback related to fraud (received items are not as purchased, sent counterfeits, fake reviews towards transactions). According to the questionnaire, their misconduct may be explained by the nature of the fraud since 18% to 29% of the respondents would risk fraud when facing the temptation of more money.

The intuition behind this unexpected experimental result is intuitive. As commonly demonstrated, blockchains are cryptographically secured ledgers, or in some senses, they are database of records. It is appreciated that once a transaction is accepted into a block and once a block is appended to the ledger, it cannot be changed or reversed. In other words, blockchains are immutable, which is a perfect record-keeping technology. This feature is regarded as the ability to guarantee secure transactions. However, it is noteworthy that blockchains cannot by themselves support a decentralized marketplace, which consists of other components of the whole ecosystem.[7] Market users transact with each other through off-chain assets or services that are out of the control of the blockchains. Contrary to popular belief, blockchains cannot confirm whether the sellers deliver the rightful items or the buyers rate the items/services rightfully. Blockchains cannot ensure that agents conduct truthfully during real-world transactions. Based on our findings and blockchains' properties, we naturally raise our first point that blockchain technology alone cannot serve as the proposed method of avoiding or at least mitigating fraudulent behaviors for peer-to-peer e-commerce. This challenge would be the critical one faced by any decentralized application in the future.[8]

---

[4] In fact, we categorize the combination of blockchain technology and e-commerce into two types. One is traditional centralized e-commerce platform with blockchain applications; for example, blockchain could be applied in tracing products and keep product information immutable. The other is an e-commerce platform with fully blockchain enabled, in which tokens are issued and the ideology of decentralization is applied. In this paper, we aim to enrich the study on the second type of e-commerce platform and investigate on the issues and concerns from the users' perspectives.

[5] It is worth noting that we are unable to collect survey from real buyer/seller who transact on blockchains-based marketplaces due to the anonymity underling blockchains.

[6] We also conduct a regression analysis to show that our survey is not driven by particular demographic factors. See Appendix A for details. We believe that to some extent our results provide insightful users' feedback on fully blockchains-based e-commerce field, although the current collected data is not adequate to generalize.

[7] See OpenBazaar (OB), a blockchain enabled peer-to-peer marketplace, is built on Hyperledger architecture. https://docs.openbazaar.org/.

At this early stage, before adopting blockchains for e-commerce, it is helpful to investigate how to protect transacting parties and enable them to seek recourse from failed performance while keeping the immunity underlying the marketplace. To contribute to this exploration, we ask students (i) who should take responsibility for providing trust and (ii) whether they feel safer if there were a reputation system linked to the users' true identity in a decentralized ecosystem. Our questionnaire survey shows that Chinese participants rely more on the government, but Norwegian participants believe in companies to provide trust. It is worth noting that linking buyer/sellers' true identity to a reputation system makes 89% of students feel safer. Such a mechanism might impede the advantage of privacy protection from blockchain. To evaluate its effects, we further ask students' privacy concerns. Our survey indicates that Chinese users have higher privacy awareness, while Norwegian students are comfortable sharing their data.

According to our survey results about detected fraud issues and users' feedback about how to promote trust in the decentralized marketplace, we attempt to propose a simple but effective conceptual model in which a central party must exist to not only resolve misconduct but also guarantee the "purity" of decentralization. Specifically, anonymous users transact on the platform as the usual blockchain-based marketplace. In the event of disputes, either or both parties could file for the trusted authority to arbitrate by providing the case detail and their true identity. The authority is entitled to the right to downgrade the fraudulent side's credit record, which is stored by a permissioned blockchain accessed only by the authority. Arguably, our mechanism allows users to enjoy anonymity and ensure that users can make the correct choices.

---

[8] In reality, OB proposed on its website that it can eliminate counterparty risk by a third-party "moderator". We provide a closer analysis below to reveal that such a mechanism cannot fully and effectively resolve the risk.

In the rest of the paper, we provide a literature review in related areas in the next section. In Section 3, we illustrate how to set up the experiments and we discuss what we found in Section 4. Last we conclude in Section 5.

## 2. Literature review

The blockchain was first introduced as a mechanism to prevent double-spending in the peer-to-peer electronic cash system known as Bitcoin. Based on the underlying idea of Nakamoto (2008), blockchain technology combined with smart contracts has outgrown its origin in cryptocurrencies and is heading to various commercial applications (Nofer et al. 2017). Casino et al. (2018) provide a systematic literature review of blockchain-based applications across diverse sectors such as supply chain, business, healthcare, IoT, privacy, and data management. Within our research scope, we focus on trust-free systems and propose to address trust issues in peer-to-peer e-commerce systems by investigating and mitigating the incidence of misconduct behaviors.

### 2.1 E-commerce

Lim et al. (2019) provide a survey of existing blockchain technologies and applications in e-commerce and highlight key blockchain properties and their benefits and challenges in online shopping sites. Sheikh et al. (2019) summarize various e-commerce industries implementing blockchain-based platforms to improve online businesses. Sherman et al. (2019) propose a model that combines the idea of a consumer-consumer (C2C) and blockchain technology to mitigate worries between consumers by giving the consumer the freedom to make decisions in dealing with the consumers of others, and by tracking their rates and feedback, depending on previous business transactions.

There exist some studies that focus on the system design in blockchain-based e-commerce, such as protecting privacy (Gallay et al., 2017; Green and Miers, 2017; Jiang et al., 2019), promoting digital payment systems (Roos et al., 2017; Kokoris-Kogias et al., 2018; Pass and Shi, 2018), enhancing traceability in the supply chain (Tian, 2016; Xu et al., 2016), and incorporating supervisions while maintaining decentralization (Chen et al., 2017; Liu et al., 2018).

The adoption of blockchain technologies in e-commerce necessitates the consideration of system reliability and resilience, as these attributes are pivotal in ensuring trust and efficiency within decentralized marketplaces. Extensive contributions to reliability modeling exist that provide foundational insights into system performance under various conditions, which are relevant to blockchain-based e-commerce systems (Cheng and Elsayed, 2015, 2017a, 2017b, 2018, 2021a, and 2021b). Related studies on resilient system design demonstrate approaches that can be adapted to blockchain systems to ensure robust transaction handling and dispute resolution, which is an aspect critical to fostering trust in blockchain networks (Cheng et al., 2021a, 2021b, 2022, 2023, and 2025; Gao et al., 2021 and 2022; Li et al., 2024; Wei at al., 2024 and 2025; Zhao et al., 2025; Zhou et al., 2022 and 2023).

Broader studies enrich the discourse on blockchain reliability and applications. Alzahrani and Bulusu (2023) discuss blockchain's role in enhancing data integrity and trust, while Singh et al. (2023) highlight its importance in building resilient supply chains. Kamble et al. (2022) explore organizational strategies for blockchain adoption, and Gao et al. (2021) propose robust design strategies applicable to e-commerce systems. Collectively, these studies provide a holistic view of blockchain's transformative potential when paired with advanced reliability and resilience frameworks.

By integrating reliability and resilience modeling with blockchain technologies, the development of secure, efficient, and fraud-resistant e-commerce platforms becomes feasible. These insights support the creation of mechanisms that address fraudulent behaviors and optimize decentralized marketplaces for broader adoption. The simulation performance of those designed platforms shows they are usable implementations; however, none provide a real test for e-commerce users in practice.

**2.2 Information fraud**

To preserve users' privacy and prevent information fraud, scholars have been focusing on re-designing reputation systems in the era of blockchain technology. For instance, Dennis and Owenson (2016) propose removing human opinions from reputation systems. Instead, the reputation is represented by a binary value, which reflects if the users receive the file. In this case, the systems contain only objective information. However, many other factors, e.g., product quality, are also very important to customers' purchasing decisions.

As privacy is an essential concern for users who are reluctant to provide information, Schaub et al. (2016) propose a fully decentralized reputation system atop a public blockchain with blind signatures to achieve consumer anonymity. Customers and sellers use private and public keys to communicate with each other. However, the openness of a public blockchain and consumer anonymity may raise the concern of Sybil's attacks.

Similarly, Soska and Christin (2015) propose an anonymous reputation system based on ring signatures and blockchain technology's robust transaction chain property. In their model, consumers are required to pay specific fees for injecting ratings. Such a preventive mechanism increases the costs of fraud so that it can mitigate Sybil's attacks. However, they are ineffective if the perceived benefit from attacks exceeds the price.

Those extant studies investigate the trust/fraud concept of blockchain technology and applications primarily from a technology-focused perspective that is closely related to engineering, programming, and computer science, evidencing that there is a need to study the topic more intensively from the user angle (Risius and Spohrer, 2017).

**2.3. Survey/Experiment**

The empirical research related to trust and Blockchain technology is still relatively scarce. Most research indicates that existing blockchain-based solutions significantly contribute to the trust mechanism.

Fleischmann and Ivens (2019) suggest that trust is a key driver for user/consumer adoption of blockchain technology and applications by using an inductive research approach that builds theory from qualitative empirical data (as in Sarker et al., 2013). The sample consists of interviews among 14 experts with extensive expertise in blockchain technology and an online survey from 19 U.S.-based Bitcoin users via the Amazon MTurk platform.

Gonzalez (2019) finds that blockchain technology is needed in peer-to-peer (P2P) lending platforms to assist in monitoring and bad loan recovery. Specifically, this study examines 909 lending decisions by 303 finance students on a mock P2P site in which either male or female loan applicants are reported to be highly trusted by other lenders. Overall, the investors who have experienced financial trauma appear more susceptible to trust-enhancing heuristics.

Milosav (2019) compares two experimental treatments conducted over Amazon MTurk and utilizes a between-subject design. The first treatment is based on a simple Trust game designed by Berg et al. (1995) without the social history report, while the second treatment aims to operationalize a blockchain-based smart contract. The results presented in the paper indicate that if Blockchain technology is implemented correctly, it might promote trust and trustworthiness in human-to-human relationships.

Schwerin (2018) attempts to address how blockchain technology, characterized by decentralization, immutability, and digitized values, will be affected by the strict privacy regulation of the European General Data Protection Regulation(GDPR) and vice versa. The questions are investigated using a Delphi study that asks a panel of 25 renowned experts to find opportunities, limitations, and general suggestions for both topics.

Wamba (2019) proposes the importance of trust among supply chain stakeholders in supporting the continued use of blockchain technologies-enabled supply chain applications, supported by the data collected among 344 supply chain professionals in India.

The above articles take a broad approach, exploring the benefits of blockchain technology and applications. However, these insights may not provide guidance tailored to specific use cases, such as identifying the most important users' concerns/thoughts about blockchain. Our work indeed provides a deeper understanding of which improvement may best promote users' acceptance of particular applications.

**2.4. Contribution**

The innovative contributions of this article are two-fold. First, this research explores the role of trust in blockchain technology and applications from the user perspective. By building a behavioral experimental study analysis, this paper takes a scarcely researched, alternate angle on the trust aspect in blockchain technology and applications, expanding and complementing extant, primarily technology-oriented and general survey research.

Second, our findings suggest promoting more extensive interdisciplinary work programs investigating how blockchains affect fraudulent transactions, trust's role, and mediators' functions. Moreover, this research paper adds to the existing body of literature by answering the following questions: Who is the trusted authority managing agents' social credit score from users' perspective regarding blockchain technology and applications?

Therefore, our results can help organizations make the best use of nascent technology in strategic and organizational decisions. This research also unveils comprehensive insights that may be critical within the technology adoption process of blockchain applications. These insights may serve researchers for future blockchain-related work and guide practitioners in developing powerful blockchain applications that have a chance of users' acceptance and that help to build trust with users.

**3. Experiment**

This section presents a detailed overview of the experiment. We first introduce the decentralized peer-to-peer marketplace used for the trading experiment. Second, we elaborate on how the experiment is implemented.

## 3.1 Experiment platform

We create a decentralized e-commerce marketplace via the platform provided by Origin Protocol, Inc.[9] For each group of experiments, a new independent marketplace will be created to initiate an empty one with no listings. This application provides a set of features that are useful for our experiment. Figure 1 displays a flowchart that includes the essential activities in the marketplace, and an overview of key features will be demonstrated.

--Insert Figure 1 here—

First, users must connect to the Ethereum network to interact with the marketplace built on Origin's platform. The Origin community recommends using the Metamask Chrome Browser Extension to run Ethereum DApps right in the browser without running a full Ethereum node.[10] Next, sellers can create a listing to sell, rent, announce, or exchange something. Active listings can be searched, browsed, and booked via a frontend DApp. At the same time, buyers can make offers to sellers for listings. Once an offer has been made, there are three possible transitions: buyer withdraws, seller declines, or seller accepts. The first two will result in the buyer being refunded. If the seller accepts, the offer becomes the virtual contract for the transaction between them. Once the buyer finalizes this transaction, the funds in escrow are transferred to the seller, and the transaction is considered complete. Finally, users are encouraged to leave feedback about the interaction in the form of a rating or review after the transaction is completed, which enables users to establish their reputations over time with verified transactions.

## 3.2 Experiment procedure

We conducted five trials, 4 in Tianjin, China, and 1 in Trondheim, Norway. All the experiment participants are students, which limits them to the group of society that is likely to be future potential users of blockchain products. High-educated university students often have above-average knowledge and understanding of digital technology. Moreover, educated targets can demonstrate objectiveness and well-reasoned subjectiveness. Therefore, students make an ideal target audience for technology-concerning experiments.

---

[9] https://www.originprotocol.com/en/creator.
[10] See generally: https://metamask.io/.

The procedure of the chronological tasks the participants will follow:

(1) By following the on-screen instructions, Each student installs Mozilla Firefox and the Metamask browser extension to create a Metamask account.

(2) Each student needs to request 0.5 test ETH funds to use in this experiment from the Rinkeby faucet.[11]

(3) After the setup is ready, everyone enters the marketplace deployed on the tested platform. In addition, a projector will display instructions on how to use the platform.

(4) All participants randomly draw five paper cards from a pile of 10 sets of 26 English alphabet cards. These cards represent the items for trading and have an unknown value during the trading hours. However, the value of the items is pre-determined before the game for the sake of fairness.

(5) When everyone is ready, the students start to trade on the marketplace following the instructions illustrated in the previous subsection. In fact, participants are allowed to do nothing. However, some hints regarding to the value of particular letters will be announced at certain stages during the trade hours, which may affect students' incentives to trade or not. These hints could be:

   (i) Some letters (e.g. A, E, I, O, U) might increase or decrease a certain amount of ETH to your portfolios.

   (ii) Sequential letters may be the most valuable combination.

   (iii) Duplicated letters will deduct a certain amount of ETH for each duplicated card.

In addition, winners will be awarded prizes, which increases students' incentive to trade.

(6) The rating score is used to describe the seller's reputation. A low rating score may result in few or no sales. One can neglect the buyer's reputation in this case because of the escrow feature provided by Ethereum's smart contract controls the payment.

(7) The value of single letter and combination of letters will be revealed to students after the

---

[11] See generally: https://faucet.rinkeby.io/.

market is closed. Note that positive (negative) value of those values will be added (deducted) from participants' portfolios in ETH.

(8) Winners receive prizes as an incentive to win the game. During and after the ongoing experiment, participants are advised to comment and discuss any matter regarding the concept.

(9) After the experiment, the participants will be asked to answer a questionnaire survey containing multiple-choice questions, which will further supplement the experiment data collection.

## 4. Results and analysis

This section shows discovered results from the questionnaire survey. As a part of the data collection methods, the questionnaire was required at the end of each experiment, and all participants have completed it. There was a total of 133 respondents and the results partition into 14 Norwegian respondents and 119 Chinese respondents. The results from China are combined, as the responses were considered equivalent from all trials. However, the results from Norway have a smaller confidence level because of fewer respondents. Regardless, the results from this survey still provide a satisfactory overview of the matters as the participants have tried a concept and possess adequate knowledge in the blockchain field. It is important to note that the current collected data is not adequate to generalize, but to some extent we try to provide a special perspective to get feedback on blockchain technology. The trading activities of each trial are listed in Table 1.

--Insert Table 1 here—

### 4.1 Fraud issues

Blockchain offers peer-to-peer digital transactions without involving intermediaries. That is, there is no central authority controlling and managing the transactions. The participants do not need to reveal their identity while transacting on blockchain. In our experiment setting, the marketplace makes for a decentralized ledger for storing, sending, or receiving listed items. Similar as current online shopping platforms, e.g. Amazon and Taobao, we also have a reputation mechanism imposing a review on the seller and buyer after the transaction. We would like to observe if the feedback system still works to prevent fraud when the feedback is immutable.

Since e-commerce buyers are not able to see or touch an item in person before buying, buyer's decision of purchase focuses on trusting the seller. For overcoming payment fraud in the experiment, the escrow function ensured a 100% secure payment operation. This feature make the platform convenient for the users to control and tract their payments. It also shifts the trust burden from buyers against third-party to smart contracts. This shift ensures that the payment is guaranteed to transit between buyers and sellers directly. However, our experiment results reveal that fraud could easily succeed since the payment escrow cannot guarantee what items delivered to buyers and what rating or reviews left by buyers.

Table 2 shows what we found from the experiment and the survey related to fraud issues in the e-commerce platform. We classify the questions into 3 categories: *Fraud in the experiment* is related to what happened during the experiment; *Fraud in nature* is related to the misconduct behavior in nature when people are facing temptation; *Thoughts on the problem* is related the potential mechanism to regulate misconducts.

--Insert Table 2 here—

We can see that fraud did happen during the experiment. 16% of participants are attempt to cheat (not happened yet) and more than 17% of subjects are cheated (happened). This implies that about 1/3 of subjects have the incentive to play the trick to others during the trading game. 18%-29% of the responders would take the risk to fraud when facing temptation of more money. Surprisingly, even if there is not temptation, there could be 2% of being dishonest on the rating.

Therefore, we could see that the rating system might not work well. This rating system is similar to what is used in the centralized e-commerce platform but with immutable characteristics. In a centralized platform, disputes related to fake ratings, fake comments and fake goods are still common. All these disputes rely on arbitration from the platform to resolve. Such centralized arbitration mechanism seems not efficient as the misconducts are costless and could easily get restarted in other platforms. More importantly, the misconduct records could be easily mutable by the centralized platform[12].

**4.2 Conceptual trust system**

Our experiment shows that even if the network maintains immutable records of transactions and ratings, fraud could happen and trust is still difficult to obtain and maintain. It leaves an open question: how can we construct a mechanism to induce honest and reduce fraud? Recall that Table 2 shows that 94% of participants agree that extra social benefit could give incentive to promote honesty, 87% agree to give punishment on misconducts by limiting social rights, and 89% of participants feel safer to use e-commerce if there was a reputation system linked to the users' true identity. Therefore, a more efficient system of providing trust cannot only reflect review or feedback like most recent e-commerce does. It might have the function to contribute to something like social credit system that could have an influence in real life (lowering tax rate as a reward or downgrading credit score).

We propose a conceptual mechanism to resolve the fraud issue and provide trust in Figure 2. In the model, we introduce a third party as a central authority to arbitrate the fraud case. Real life credit records are stored by a permissioned blockchain controlled by the central authority. At first, anonymous users (normally two users involved) in the blockchain based platform have disputes on transactions, e.g. fake goods. Either or both users could file for an intermediary arbitration by providing the case details as well as their private information including their real identity, trading information etc. In other words, there is no anonymity in the arbitration. The authority is entitled the rights to impose a punishment/award on real life (e.g. put records on a credit record system) and impose platform treatment (e.g. cancel the transaction or refund).

--Insert Figure 2 here—

---

[12] See one example from http://news.cntv.cn/20120813/105288.shtml

At present, OpenBazzar manages the counterparty risks of online trade with Bitcoin using multisignature escrow transactions. The buyer, seller, and a third party "moderator" create an address that requires 2-of-3 signatures to release funds.[13] If the buyer or seller have problems with the transaction, they can initiate the dispute resolution process. The moderator investigates the situation, and co-signs with the winning party to release the funds. Such a preventive strategy in fact cannot eliminate counterparty risk as it proposed. Technically, anyone can become a moderator or, given that anyone can create multiple pseudonymous accounts, pretend to be multiple moderator. Three points bear emphasizing. First, pseudo-moderator can collude with buyer/seller. Second, our survey indicates that users are willing to trust government or companies rather than individuals. Lastly but the most important one is that the buyer, the seller, and the moderator's identifies are anonymous during the process of arbitration, which may make OB's resolution mechanism is not as effective as propose.[14]

It is re-emphasized that the authority gets access to true identities of buyer/seller in the arbitration and is able to punish default party by putting records into his/her social credit. This feature makes the cost of misconducts far outweigh any benefits from fraud. In other words, our hypothetical system is similar to legislation system in the real world. It has cost to file a lawsuit and there would be an immutable crime record if found guilty. We argue that fraud rate should be reduced if this system is introduced into blockchain based platform. However, who should be the central authority to provide trust and how does users think about giving up private information are two remaining questions to build up such system. We will discuss them respectively in the next steps.

**4.3 Who is trusted to provide trust?**

Figure 3 shows the percentage of beliefs on who should have the responsibility of providing trust in decentralized blockchain e-commerce. It is clear to see that the Chinese rely more on the government, while half of the Norwegians think companies are responsible. The results also reflect the type of market economy the respective countries have. In China, many of the governmental decisions are made based on the state's ideology, philosophy and interest. For instance, the existence of the Great Firewall of China divides in social, political and economic reasons. People believe that government has more power on network development and thus it should be safer to have government's support and guarantee on online transactions. For the Norwegian companies, there are some companies which are partly state-owned and regulated, but the majority of companies are governed privately in a free market economy. Legislation is matured in well-developed markets which ensures more trust in civilian culture. Thus, it seems taken for granted that the government in China are responsible for ensuring a system of trust in e-commerce. Likewise, the companies have that same responsibility in Norway.

---

[13] See generally: https://openbazaar.zendesk.com/hc/en-us/sections/115000725872-Moderators.

[14] Recently, OB adds "verified moderators" to its dispute resolution mechanism. The verification requirements are complex and require candidates to be technology literate. However, their real identity is still anonymous.

--Insert Figure 3 here—

The trust system built by the authority is useless without quality-assured input data. It requires cooperation from the users to provide the private information. Is there any concern from the users to give up private information to the entity?

**4.4 Privacy issue**

Figure 4 reports the willingness to share private personal data in exchange with social safety and security. Interestingly, 59.7% of Chinese students are not willing to undertake this type of exchange. Basically, Chinese people might not want to share the personal data but was forced to do so due to centralization purpose. However, many companies and government systems have security weaknesses, which often leads to abuse of personal data and leaks of personal data. As a consequence, the majority of Chinese people are generally afraid of sharing their personal information.

--Insert Figure 4 here—

In Norway, the high percentage speaks for itself. Norway has one of the world's most advanced and comprehensive legal system. Besides, citizens' privacy is well protected by laws. Thus, the Norwegians have more control of their personal data with fewer cases of identity fraud. The Norwegian participants could, therefore, share a higher willingness to exchanging their personal data with social safety and security.

Figure 5 compares the willingness to share all transaction data to the governmental institutions to impact the participants' credit line. We focus on credit line incentive as students have financing constraints and loan is popular in e-commerce for students. As displayed, 84.4% of the Chinese participants would agree to share transaction information to improve their creditworthiness. On the contrary, 64.3% of the Norwegian participants disagree in sharing such information. One explanation could be that it is tough to obtain a loan and receive approval on credit applications for Chinese students. On the other hand, Norway has an advanced welfare system with ease for students to apply for student loans and credit.

--Insert Figure 5 here--

The other reason is that transaction data is not that 'private' or important since online shopping is popular in China. Among all the participants, more than 42% of the Chinese students are buying items online more than one time during a week. Contrary, all the Norwegian students are purchasing items online either one time in a month or only a few times in a year. These numbers indicate a more frequent online shopping habit among Chinese participants.

Both Figure 4 and Figure 5 describe different motives to exchange the willingness of sharing private data. No matter if it is personal data or transaction data, they are valuable and useful for companies and the government. By comparing Norway with China in the sense of data privacy, it is clear to distinguish a fundamental difference in the practice of collecting, storing and sharing data. Chinese internet companies have in the past years built business models around Chinese people's lack of awareness about privacy. It is also statutory to share these data with the government to avoid revoking their business license. Besides, numerous data leaks incidents as the consequence of security malpractice are simple targets for fraudsters and criminals. The China Consumer Association[15] reported that 85.2% of their 5458 survey respondents had experienced personal information disclosure such as leaked phone numbers and ID information. These series of privacy breaches were the needed wake-up call for the privacy unawareness in the population.

Today, Chinese users have higher privacy awareness as a result of facing the consequences of data abuse. Especially the younger generation who carries a more profound knowledge in digital technologies. From the questionnaire in the experiment, more than 80% of the participants would not sell their privacy for a monetary value. Despite the massive amount of mobile applications and data collecting devices in the Chinese market, most people do not coincide with the statement from the CEO of Baidu Yanhong Li[16] *"the Chinese people are willing to trade privacy for convenience, safety and efficiency"*. As also observed from the experiments and survey, the students were cautious of every keystroke and did not enter any sensitive information. In all, the importance of privacy in people's life has indulged in a government intervention of building a data protection framework.

Similar to what is happening in China, the importance of privacy is fundamental in Norway. The level of privacy awareness rises accordingly with the expansion of digital services and devices. With a previously steady legal system in protecting privacy, it seems like the Norwegian students are comfortable in sharing their data and trust different services. Comparing to the Chinese students, the Norwegian students had a laxer attitude when entering data. Despite the high concern on privacy, this attitude reflects good faith in society and the ease of information sharing in a well-protected system.

In all, from the survey we can imply that privacy is a big concern but still it could be compromised for a better life. According to the survey, over 80% of the participants agreed that decentralization in this experiment made them more confident trading than having an intermediary controlling the trade. Thus, the potential of blockchain technology seems to be compliant with people's privacy expectations. However, perfect anonymity also brings potential fraud issues. To impose higher cost on fraudulent users, partial anonymity could be an option. That is, we could maintain anonymity in normal transactions but break it by giving up privacy in fraud cases.

---

[15] https://www.docin.com/p-2154704096.html
[16] https://www.gamersky.com/news/201803/1029448.shtml

## 5. Conclusion

In this paper, we verity whether the blockchain can serve as the technology underlying decentralized marketplaces to promote trust. We first invite several groups of students to transact on a decentralized peer-to-peer marketplace built on the platform provided by Origin Protocol Inc. and then we conduct the survey about their experience of usage. Based on our survey results we find that 33% of respondents play tricks to others. We next propose a conceptual mechanism that aims to mitigate fraudulent behaviors according to respondents' thought about who should take responsibility for providing trust and at the same time protecting their privacy. Therefore, blockchain is not ready to take on all of the challenges as a payment technology, as we are still trying to figure out where in the payment ecosystem it belongs. But we do look forward to new developments with blockchain and believe that it will play a role in securing ecommerce technology in the future.

**Appendix A. Demographic effects**

We have discussed some concerns related to blockchain based decentralized e-commerce in a general manner by taking all responders as a whole. In this subsection, we go deeper and see if specific demographic factors are associated with the innovation from e-commerce. In our experiment, we have detail personal information for most of the participants in Chinese group. We then investigate the demographic effects via regression analyses from this subsample.

Table 3 shows the summary statistics and correlation of the variables of interest and demographic variables. In this table, we focus on a subsample from Chinese participants when their demographic information is available. We have 75 observations in total. Panel A shows the summary statistics and Panel B shows the correlation for all the variables in regressions.

--Insert Table 3 here—

We have four variables of interest. The first one is *first mover*, which is a dummy variable indicates participants' future interest in using blockchain and cryptocurrency related products. We find 52% the participants are willing to try and accept these new things. Since our participants are all students in the university, they are believed to have more acceptance for innovations than common people. It is likely that this ratio could be lower than 50% for the public in China, which implies that Chinese people are a bit conservative to new things. The second variable *Prefer Blockchain Platform* is to indicate a preference on blockchain applications to centralized platforms. Every transaction in this experiment is listed on a public blockchain (everyone can see the transactions). 85% of the participants feel more confident trading in blockchain based platforms. This implies that students hold stronger trust in today's new digital technology. We use the third variable *Trust in Government* to indicate that belief in government takes responsibility for decentralization. Over 60% expects the government to take the next step on blockchain in future. Our last variable of interest is *Privacy Concern*, which indicates unwillingness to give up private information for other reasons (e.g. money). We have more than 80% of the participants are not willing to make a deal on selling their private information.

There does not exist odd distribution of demographic variables in our sample. On average, the age is around 24 since they are all graduate students with diversified undergraduate majors. 35% of the sample are finance students and 13% of the sample may have finance background due to parents' job in finance industry. We care about finance background because the market place is trading via one cryptocurrency, i.e. ETH. So having finance background might have more exposure on this new financial tool as well as related blockchain technics, which might bias up their answers in the questionnaire. 40% of participants are male and 69% are in a single-child family. One third of the sample comes from high income family that is above median yearly level, i.e. 200,000-300,000 CNY per year.

The overall correlation among the variables is low. We then perform OLS regressions for our variables of interest on demographic variables in Table 4. We also control for province fixed effects to avoid the concern related to district omitted factors. Interestingly, we find that participants who is the single child are significantly associated with higher potential willingness to try new digital technology. This is consistent with the finding of Poston and Yu (1985) that suggests a difference between single-child and multiple-child on lifestyle. We also find that male participants seem to have less concern on privacy issues, but the significance is weak in power.

--Insert Table 4 here—

Basically, we do not find any other demographic factors that are significantly dominant in all the results. This may suggest that demographic factors are not alternative explanations to our abovementioned discussions.

**Appendix B. Variables definition**

| | |
|---|---|
| First Mover | It is a dummy variable equal to 1 if the participant is willing to try blockchain products in future and 0 otherwise. |
| Blockchain Platform | It is a dummy variable equal to 1 if the participant prefers blockchain base system to centralized system and 0 otherwise. |
| Trust in Government | It is a dummy variable equal to 1 if the participant trusts the government in providing safety/responsibility in blockchain related applications and 0 otherwise. |
| Privacy Concern | It is a dummy variable equal to 1 if the participant cares more about his/her personal information than money and 0 otherwise. |
| High Income | It is a dummy variable equal to 1 if the participant's yearly family income is above median level (i.e. 200,000-300,000 CNY) and 0 otherwise. |
| Age | It is the age of participant. |
| Male | It is a dummy variable equal to 1 if the participant is a male and 0 otherwise. |
| Single Child | It is a dummy variable equal to 1 if the participant is a single child in the family and 0 otherwise. |
| Major in Finance | It is a dummy variable equal to 1 if the participant's undergraduate is in finance major and 0 otherwise. |
| Parents in Finance | It is a dummy variable equal to 1 if the participant's parents are either working in finance industry and 0 otherwise. |

**Figure 1. Flow chart of marketplace**

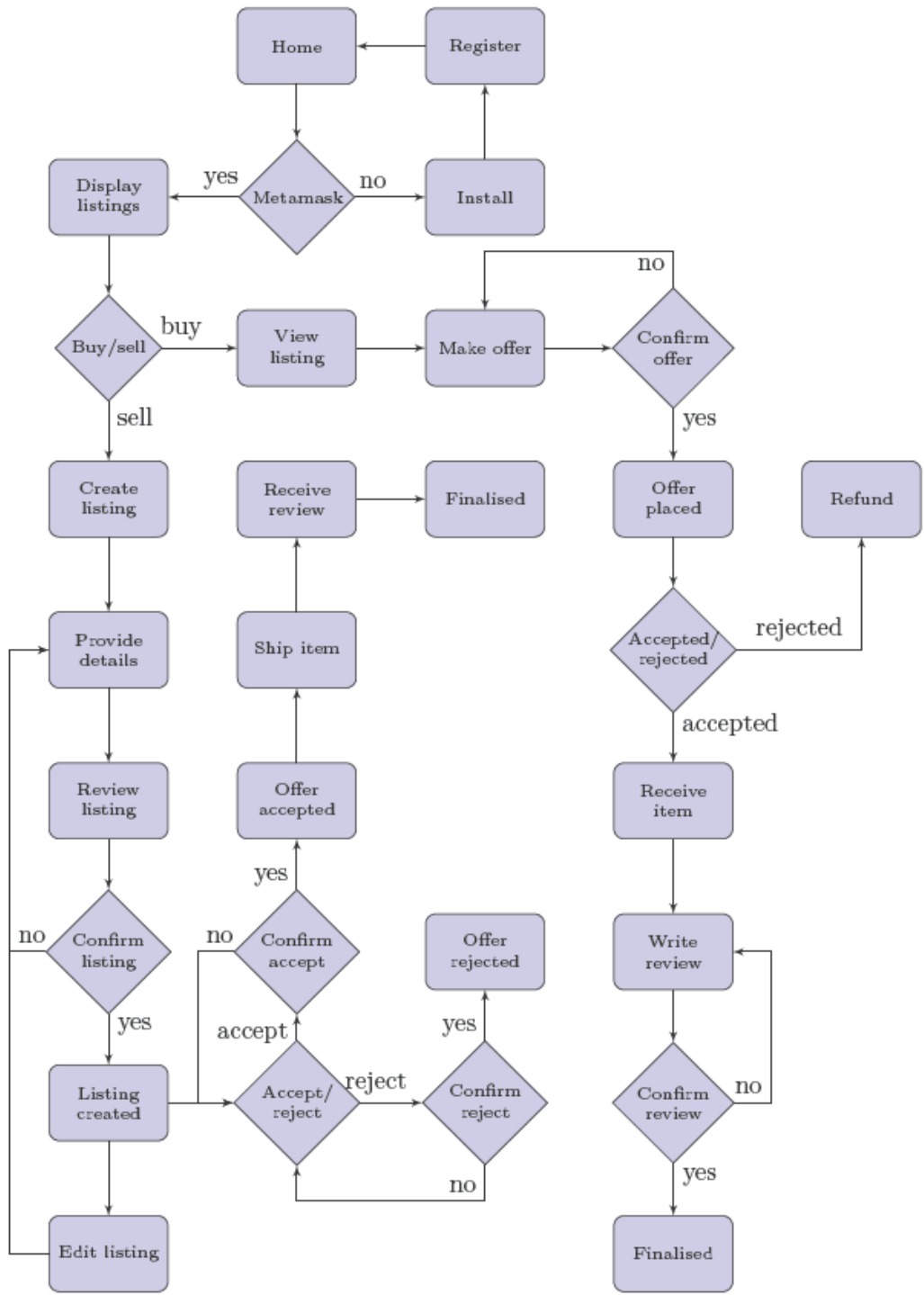

**Figure 2. Conceptual intermediary model for resolving fraud issue**

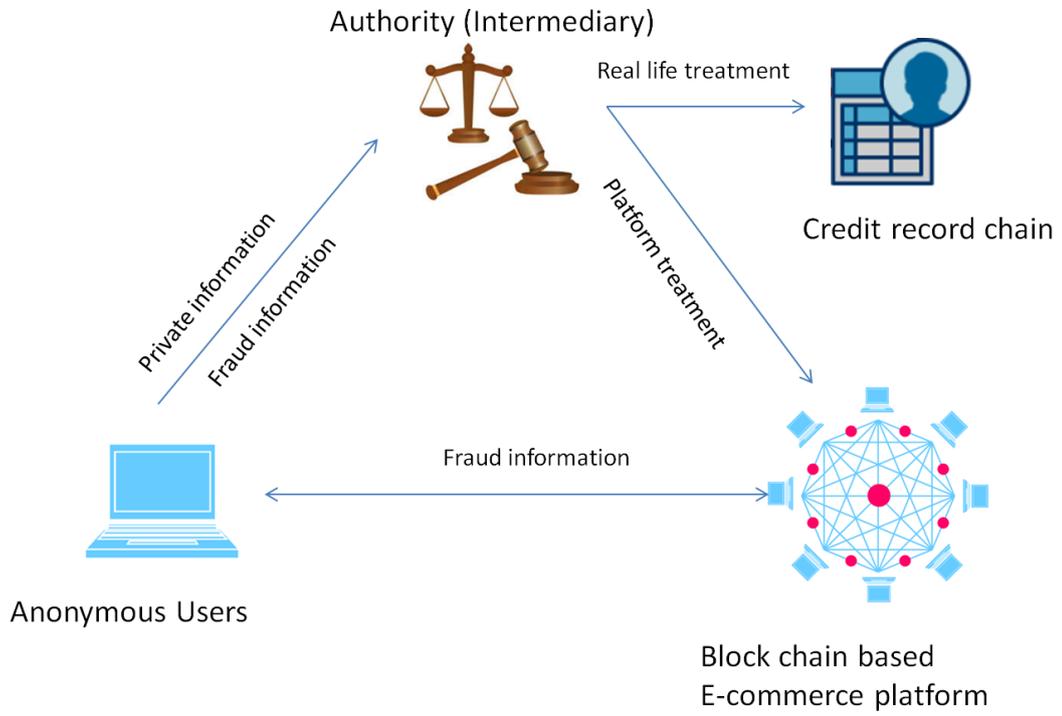

**Figure 3. Who should take responsibility for providing trust in decentralized applications in your country?**

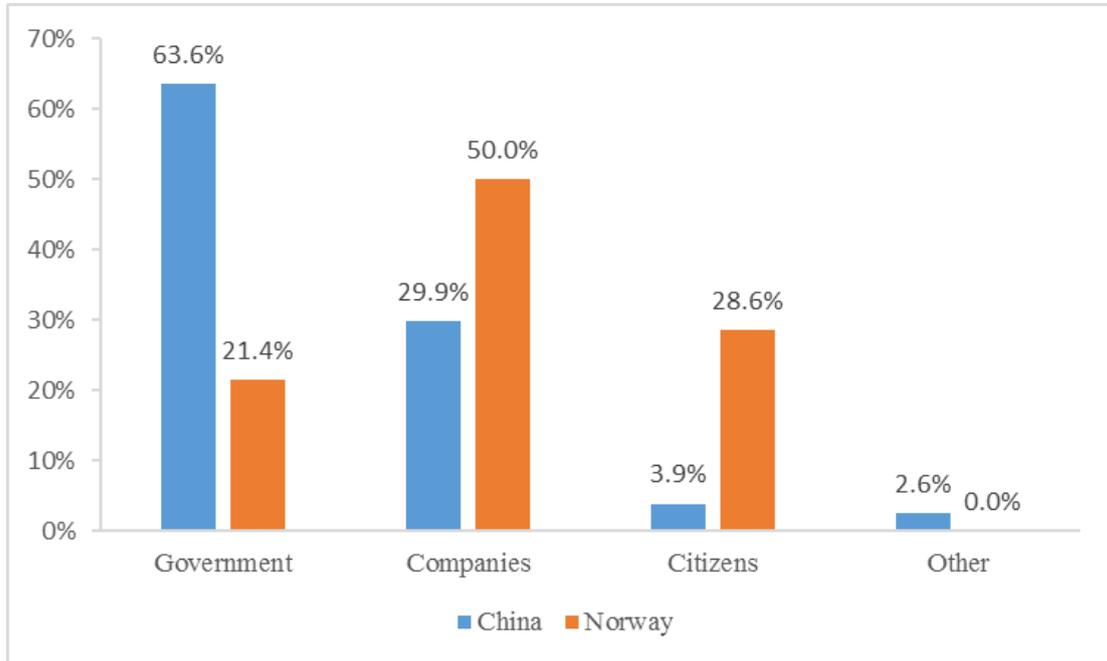

**Figure 4. Would you share your private personal data in order to be provided with social safety and security?**

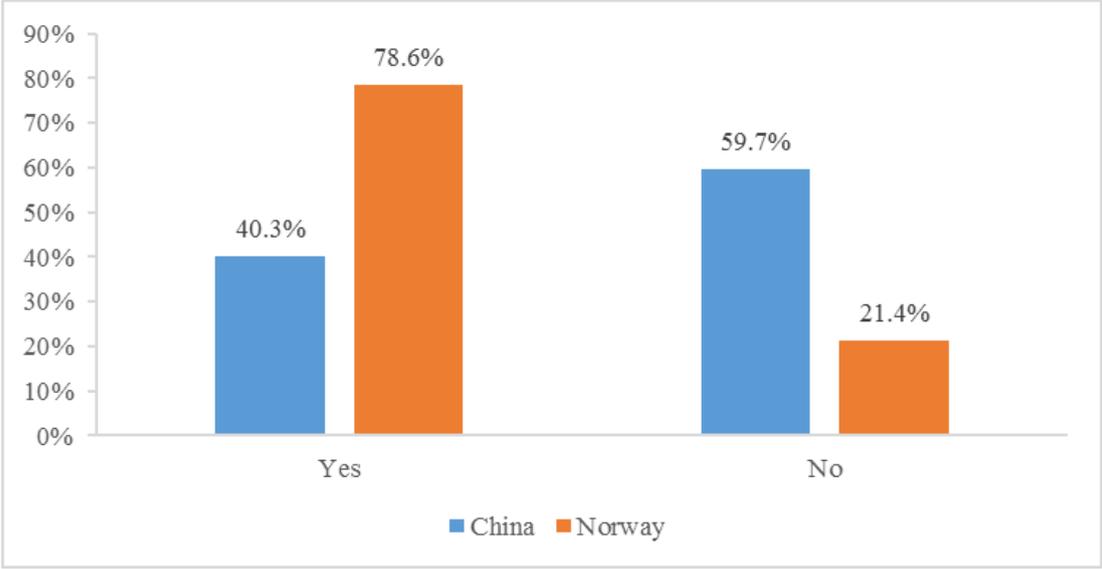

**Figure 5. Would you share all transaction data to the governmental institutions in order to influence your creditworthiness? This can ease your loan and credit applications.**

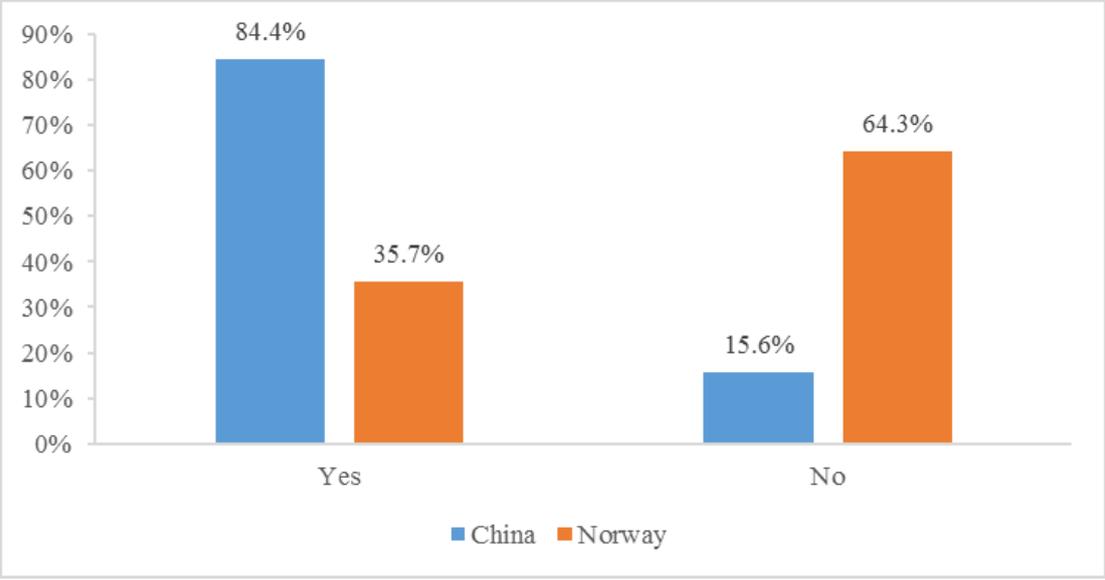

**Table 1. Summary statistics of trading activities in each trial**

This table presents a summary of the statistics of our survey data. Panel A shows the attendance and trading activities in each trial.

|  | Trial 1 | Trial 2 | Trial 3 | Trial 4 | Trial 5 |
|---|---|---|---|---|---|
| # of Participants | 78 | 16 | 13 | 12 | 14 |
| Total listings | 116 | 73 | 65 | 54 | 93 |
| Complete transactions | 35 | 47 | 32 | 30 | 37 |
| Pending transactions | 16 | 2 | 2 | 2 | 2 |
| # listings per participant | 1.49 | 4.56 | 5 | 4.5 | 6.64 |
| # completed transactions per participant | 0.45 | 2.94 | 2.46 | 2.5 | 2.64 |
| Sales per listing | 30.17% | 64.38% | 49.23% | 55.56% | 39.78% |

### Table 2. Fraud issues

This table shows the feedback related to fraud from the questionnaire. This table includes all the participants in the experiment. We classify the questions into 3 categories: Fraud in the experiment is related to what happened during the experiment; Fraud in nature is related to the misconduct behavior in nature when people are facing temptation; Thoughts on the problem are related to the potential ideas to regulate misconducts.

| Types of questions | Description | N | Mean | SD | T-value |
|---|---|---|---|---|---|
| Fraud in the experiment | Get tricked (the number of times the received goods are not as purchased) | 133 | 0.17 | 0.38 | 5.25 |
| | Attempt to trick (the number of times the sent goods are not as sold) | 133 | 0.16 | 0.37 | 4.97 |
| | Dishonest when giving a rating to others | 133 | 0.02 | 0.15 | 1.75 |
| Fraud in nature | Would you risk to fraud other people, if you had a 50% chance of becoming a millionaire without any negative consequences? | 133 | 0.24 | 0.43 | 6.47 |
| | Get tempted to fraud in order to earn more money? | 133 | 0.29 | 0.45 | 7.27 |
| | If you had access to your neighbor's computer, could you consider transfer his/her balance into your own wallet? | 133 | 0.18 | 0.39 | 5.39 |
| Thoughts on the problem | Would rewards like extra social benefit(e.g. Lower tax) make you act more honest? | 133 | 0.94 | 0.24 | 45.41 |
| | Do you find it acceptable to be punished by limiting your rights for social services if you fraud other people? | 133 | 0.87 | 0.34 | 30.01 |
| | Would you feel safer to use e-commerce if there was a reputation system linked to the users' true identity? | 133 | 0.89 | 0.32 | 32.22 |

**Table 3. Summary statistics and correlation of variables in regressions**

This table focuses on a subsample from Chinese participants when their demographic information is available. Panel A shows the summary statistics, and Panel B shows the correlation between all the variables in the regressions.

Panel A Summary statistics of Chinese participants with detailed demographic information

| Variable | Obs | Mean | Std. Dev. | Min | Max |
| --- | --- | --- | --- | --- | --- |
| First Mover | 75 | 0.52 | 0.50 | 0 | 1 |
| Prefer Blockchain Platform | 75 | 0.85 | 0.36 | 0 | 1 |
| Trust in Government | 75 | 0.63 | 0.49 | 0 | 1 |
| Privacy Concern | 75 | 0.83 | 0.38 | 0 | 1 |
| High Income | 75 | 0.31 | 0.46 | 0 | 1 |
| Age | 75 | 23.83 | 1.41 | 21 | 28 |
| Male | 75 | 0.40 | 0.49 | 0 | 1 |
| Single Child | 75 | 0.69 | 0.46 | 0 | 1 |
| Major in Finance | 75 | 0.35 | 0.48 | 0 | 1 |
| Parents in Finance | 75 | 0.13 | 0.34 | 0 | 1 |

Panel B Correlation of all variables in regression analyses

| | First Mover | Blockchain Platform | Trust in Government | Privacy Concern | High Income | Age | Male | Single Child | Major in Finance | Parents in Finance |
|---|---|---|---|---|---|---|---|---|---|---|
| First Mover | 1 | | | | | | | | | |
| Blockchain Platform | 0.28 | 1 | | | | | | | | |
| Trust in Government | -0.02 | -0.09 | 1 | | | | | | | |
| Privacy Concern | -0.02 | -0.09 | -0.06 | 1 | | | | | | |
| High Income | -0.06 | 0.19 | -0.08 | -0.08 | 1 | | | | | |
| Age | -0.04 | -0.08 | -0.17 | -0.06 | -0.02 | 1 | | | | |
| Male | -0.09 | -0.12 | 0.01 | -0.27 | -0.01 | 0.32 | 1 | | | |
| Single Child | 0.17 | -0.03 | 0.14 | 0.00 | -0.06 | -0.12 | 0.07 | 1 | | |
| Major in Finance | 0.03 | 0.06 | 0.10 | 0.11 | -0.12 | -0.11 | -0.31 | -0.06 | 1 | |
| Parents in Finance | -0.09 | 0.05 | -0.02 | 0.08 | 0.16 | -0.15 | 0.16 | -0.08 | -0.12 | 1 |

**Table 4. Demographic effects**

This table presents the effect from demographic factors on concerns related to blockchain application in the questionnaire. We focus on the subsample from Chinese participants when their demographic information is available. The definitions of variables are in Appendix B. We use ***,**,* to indicate significance at 1%, 5%, and 10% respectively.

| VARIABLES | (1)<br>First Mover | (2)<br>Blockchain Platform | (3)<br>Trust in Government | (4)<br>Privacy Concern |
|---|---|---|---|---|
| High Income | 0.183 | 0.165 | -0.056 | -0.150 |
|  | (1.17) | (1.51) | (-0.36) | (-1.29) |
| Age | 0.029 | -0.008 | -0.106 | 0.041 |
|  | (0.46) | (-0.17) | (-1.65) | (0.85) |
| Male | -0.105 | 0.013 | 0.110 | -0.232* |
|  | (-0.61) | (0.11) | (0.64) | (-1.82) |
| Single Child | 0.410** | 0.064 | 0.084 | 0.077 |
|  | (2.54) | (0.56) | (0.52) | (0.64) |
| Major in Finance | -0.005 | 0.117 | 0.071 | 0.027 |
|  | (-0.03) | (1.05) | (0.45) | (0.23) |
| Parents in Finance | 0.004 | 0.057 | 0.029 | 0.200 |
|  | (0.02) | (0.35) | (0.13) | (1.17) |
| Constant | -0.393 | 1.021 | 3.253** | -0.422 |
|  | (-0.25) | (0.91) | (2.03) | (-0.36) |
| Fixed effect | Province | Province | Province | Province |
| Observations | 75 | 75 | 75 | 75 |
| R-squared | 0.363 | 0.378 | 0.323 | 0.391 |